\documentclass[11pt,amsmath,amssymb,nofootinbib]{revtex4}
\usepackage{dcolumn}
\usepackage{bm}
\begin{document}
\title{Black hole entropy, log corrections and quantum ergosphere}
\author{Michele Arzano}
\email{arzano@physics.unc.edu}
\affiliation{Department of Physics and Astronomy\\
University of North Carolina\\
Chapel Hill, North Carolina 27599-3255, USA}
\begin{abstract}
\begin{center}
{\bf Abstract}
\end{center}
Quantum-gravity corrections to the probability of emission of a particle from a black hole in the Parikh-Wilczek tunneling framework are studied. We consider the effects of zero-point quantum fluctuations of the metric on the emission probability for a tunneling shell.   Quantum properties of the geometry are responsible for the formation of a ``quantum egosphere"  whose effects on the emission probability can be related to the emergence of a logarithmic correction to the Bekenstein-Hawking entropy-area formula.
\end{abstract}
\maketitle
\section{Introduction}
The linear entropy-area relation for a black hole, originally proposed by Bekenstein in \cite{Bekenstein:1971hc} through a series of intuitive arguments, can be viewed as a first step towards a description of  the quantum behavior of black holes. In fact the main argument presented in \cite{Bekenstein:1971hc} relies on the inclusion of a quantum localization limit for a particle crossing the event horizon when estimating the minimal increase in the black hole area during an absorption process. The localization limit, due to its quantum nature, provides the factor $\hbar$ needed on dimensional grounds to relate the (dimensionless) entropy to the black hole area.  Hawking's discovery 
\cite{Hawking:1974sw} that the evolution of quantum fields on a collapsing geometry does indeed predict a thermal flux of particles {\it away} from the horizon confirmed that the black hole entropy/area can be considered a thermodynamic quantity and it is legitimate to define a temperature that corresponds to the ``physical" temperature associated with the radiation.\\
It is interesting to note how the inclusion of quantum effects allows, for particles in a black hole geometry, to propagate through classically forbidden regions.
This suggests that it should be possible to describe the black hole emission process, in a semiclassical fashion, as quantum tunneling.
Parikh and Wilczek \cite{Parikh:1999mf} (see also \cite{Parikh:2004ih, Parikh:2004rh})  showed how such a description of black hole radiance is possible if one considers the emission as a transition between states with the same energy.  In this way the lowering of the mass of the black hole during the process and the related change in the radius set the barrier through which the particle tunnels.  The resulting probability of emission differs from the standard Boltzmann factor by a corrective term which depends on the ratio of the energy of the emitted particle and the mass of the hole.  The appearance of the correction causes the emission spectrum to be non-thermal. This reflects the fact that in order to describe transitions in which the energy of the emitted particle-black hole system does not change one must take into account  the particle's self-gravitation.  In the limit when the energy of the emitted particle is small compared with the mass of the  black hole the emission spectrum becomes thermal and  Hawking's result is recovered.\\
In the tunneling picture the  Bekenstein-Hawking entropy-area relation can be deduced from the form of the emission probability.  In fact for a generic system undergoing a quantum transition the emission probability  is proportional \cite{Parikh:2004ih, Parikh:2004rh} to a phase space factor depending on the initial and final entropy of the system.  A phase space factor given by the exponential of the difference between the Bekenstein-Hawking entropy $S_{BH}=\frac{A}{4}=4 \pi M^2$ associated with the black hole after and before the emission corresponds exactly to the Parikh-Wilczek result for the tunneling probability\footnote{The same result for the emission probability is obtained, using different techniques, in \cite{Massar}.  In the same work the authors discuss the universal validity of the formula $\Gamma \sim e^{\left(-\Delta A/4\right)}$ for a quantum emission from {\it every} type of event horizon.}. \\
The derivation of Parikh and Wilczek gives a dynamical description of black hole radiance in terms of the semiclassical tunneling of a shell propagating in a Schwarzschild metric. The metric ``knows" of the particle's energy through the phenomenon of back reaction but its role is just that of a {\it classical} background space-time.
It is interesting to ask then if it is possible to have a complementary derivation of black hole radiance in which space-time itself with its ``quantum" properties drives the emission process.
In \cite{York:1983zb} York provided such a description in terms of zero point quantum fluctuations of the black hole metric.  In the model he proposed such fluctuations, governed by the uncertainty principle, are responsible for the appearance of a ``quantum ergosphere".  If one associates the irreducible mass of the quantum ergosphere to the mean thermal energy of a Planckian oscillator at a given temperature the result is that, for the lowest modes of oscillation, the temperature of the heat bath is approximatively given by Hawking's formula.\\
In this Letter we show how, within the tunneling framework, the presence of a quantum ergosphere can be related to the appearance of a logarithmic correction to the Bekenstein-Hawking entropy-area relation of the type emerging in different quantum gravity scenarios  \cite{Rovelli:1996dv, Ashtekar:1997yu, Kaul:2000kf, Strominger:1996sh, Solodukhin:1997yy, Carlip:2000nv, Govindarajan:2001ee, Gour:2002pj}.  This provides a link between quantum gravity microscopic description of black holes and the origin of the quantum fluctuations responsible for the formation of the quantum ergosphere.\\
In the next section we start with a brief review of the standard tunneling argument.  In section III we discuss the motivations which lead to the introduction of a ``quantum ergosphere" and its role in York's model for black hole radiance.  In section IV a modification of the tunneling picture is proposed in which quantum effects of the geometry, through the appearance of a quantum ergosphere, are included. It is shown how modifications of the emission spectrum due to such effects support a entropy-area relation with a leading order log-area correction. The closing section V provides a brief discussion of our results. 
\section{Tunneling through the horizon}
Following \cite{Parikh:1999mf} we obtain here an expression for the tunneling probability of a spherical shell through the horizon of a Schwarzschild black hole.  The two main ingredients of \cite{Parikh:1999mf} are the use of the WKB approximation for the tunneling probability and an effective action describing the system which includes the shell's self-gravitation.  The first approximation is valid since wave packets propagating from near the horizon are arbitrarily blue-shifted there, the geometrical optics limit applies and we can treat the shell as a particle.
In the WKB approximation the tunneling probability is a function of the imaginary part of the action
\begin{equation}
\label{tunnelampl}
\Gamma\sim e^{-2\,\mathrm{Im}\, S}\,\, .
\end{equation}
The explicit form of the action needed to compute the emission probability can be found in \cite{Kraus:1994by}.  There the corrections to the geodesic motion of a spherical shell due to self-gravitation in a Schwarzschild geometry were calculated and their consequences for the Hawking radiation spectrum were studied 
(see also \cite{Keski-Vakkuri:1996xp}).
One starts by considering the metric for a general spherically symmetric system in ADM form 
\begin{equation}
ds^2=-N_t(t,r)^2dt^2+L(t,r)^2[dr+N_r(t,r)dt]^2+R(t,r)^2d\Omega^2\,  .
\label{metricADM}
\end{equation}
Once the action for the hole-shell system has been written in Hamiltonian form, the dependence from all the momenta, but the one conjugate to the shell radius, can be eliminated using the constraints of the theory.  Integrating over the gravitational degrees of freedom and fixing the gauge appropriately 
($L=1$ $R=r$) \footnote{This choice of the gauge corresponds to a
particular set of coordinates for the line element (Painleve' coordinates) which is particularly useful to study across horizon phenomena being non-singular at the horizon and having Euclidean constant time slices (for more details see \cite{Kraus:1994fh}).} one obtains the following effective action for a massless self-gravitating spherical shell 
\begin{equation}
S=\int dt \left(p_c\dot{\hat{r}}-M_+\right)\, .
\end{equation}
Here $p_c$ is the momentum canonically conjugate to the radial position of the shell, and
$M_+$ is the total mass of the shell-hole system which plays the role of the Hamiltonian.
In terms of the black hole mass $M$ and the shell energy $E$ we have $M_+=M+E$.
Details of the lengthy derivation can be found in \cite{Kraus:1994by}.
The trajectories which extremize this action are the null geodesics of the metric
\begin{equation}
ds^2=-[N_t(r; M+E)dt]^2+[dr+N_r(r; M+E)dt]^2+r^2d\Omega^2\,  ,
\label{linel}
\end{equation}
for which
\begin{equation}
\frac{dr}{dt}= N_t(r; M+E)-N_r(r; M+E)\, .
\label{geodesic1}
\end{equation}
An explicit form for the line element (\ref{linel}) can be obtained from the expressions of $N_t$ and $N_r$ given by the constraint equations \cite{Kraus:1994by}
\begin{equation}
N_t=\pm1\,\,\,;\,\,\,  N_r=\pm\sqrt{\frac{2M_+}{r}}\,\,.
\end{equation}
In \cite{Parikh:1999mf} 
the total mass of the system is kept fixed while the hole mass is allowed to vary. This means that 
the mass parameter $M_+$ is now $M_+=M-E$.
One then has the following expression for a radial null geodesic
\begin{equation}
\dot{r}=\pm1-\sqrt{\frac{2(M-E)}{r}}\, .
\label{geodesic2}
\end{equation}
Now consider the emission of an outgoing spherical shell for which
\begin{equation}
\label{shellaction}
\mathrm{Im}\, S=\mathrm{Im}\int_{r_{in}}^{r_{fin}}p_r dr\, .
\end{equation}
$r_{in}$ and $r_{fin}$ are radial positions just inside and outside the barrier through which the particle is tunneling. 
To calculate $\mathrm{Im}\, S$ we can use Hamilton's equation, $\dot{r}=\frac{\partial H}{\partial p}$ \cite{Parikh:1999mf},
\begin{equation}
\mathrm{Im}\, S=\mathrm{Im}\int_{r_{in}}^{r_{fin}}p_r dr=
\mathrm{Im}\int_{r_{in}}^{r_{fin}}\int_M^{M-E}\frac{dH'}{\dot{r}} dr\, .
\end{equation}
The Hamiltonian is $H'=M-E'$, so the imaginary part of the action reads
\begin{equation}
\mathrm{Im}\, S=-\mathrm{Im}\int_{r_{in}}^{r_{fin}}\int_0^E\frac{dE'}{\dot{r}} dr\, .
\label{imsf}
\end{equation}
Using (\ref{geodesic2}) and integrating first over $r$ one easily obtains
\begin{equation}
\Gamma\sim \exp\left(-8\pi M E\left(1-\frac{E}{2M}\right)\right)\, ,
\label{proba}
\end{equation}
which, provided the usual Bekenstein-Hawking formula $S_{BH}=A/4=4\pi M^2$ is valid, corresponds to 
\begin{equation}
\label{Gammaf}
\Gamma\sim \exp\left[S_{BH}(M-E)-S_{BH}(M)\right]\, .
\end{equation}
If one integrates (\ref{imsf}) first over the energies it is easily seen \cite{Parikh:1999mf} that in order
to get (\ref{proba}) we must have $r_{in}=M$ and $r_{out}=M-E$.  So according to what one would expect from energy conservation, the tunneling barrier is set by the shrinking of the black hole horizon with a change in the radius related to the energy of the emitted particle itself.\\
\section{The quantum ergosphere}
In the previous section a key step toward the tunneling description was the inclusion of back reaction effects for the propagation shell at the classical level.  The origin of the ``quantum ergosphere"  can be also traced back to a calculation of back reaction effects.  In this case one studies the response of the metric to the energy momentum tensor associated with the quantum fluctuations near the horizon responsible for the black hole emission process.  An estimate \cite{Bardeen, York:1983fx} of this effect can be given in terms of the black hole luminosity, which for a Hawking flux is given by $L_H=\frac{B}{M^2}$, with $B$ a barrier factor depending on the grey body absorption and the radiated species.\\
The quantum-induced energy leakage from the black hole \cite{York:1983fx} produces a splitting between the timelike limit surface (TLS) (on which $\dot{r}=0$ for radial null geodesics, with $r$ the circumferential radius) and the event horizon (EH), approximatively identified \cite{York:1983fx} with the locus of  ``unaccelerated" ($\ddot{r}=0$) photons.  This splitting, which is essentially a back reaction effect,  leads to the creation of a quantum ergosphere associated with the geometrically well defined difference of areas  $\delta A_{QE}=A_{TLS}-A_{EH}$.  The important point to note is that if one considers the explicit form of $\delta A_{QE}$ it is easy to see that this does not go to zero when $L_H\rightarrow 0$ (and consequently when the Hawking temperature $T_H\rightarrow 0$) as it would be expected. This reveals an intrinsic ``quantum" nature of the ergosphere and indeed it turns out that $\delta A_{QE}$ goes to zero only in the limit  $\hbar\rightarrow 0$, in which case one recovers the classical Schwarzschild structure.  This fact suggests that 
for quantum black holes, zero point fluctuations of the metric might play an active role in near horizon phenomena, {\it in primis} in the Hawking effect.\\  The above arguments served as a starting point for York's description of black hole radiance.  In \cite{York:1983zb} he proposes a model of fluctuating metric whose oscillation amplitudes are determined by the uncertainty principle.  A quantum ergosphere is formed for each mode of oscillation with an irreducible mass defined by the difference between the mean irreducible masses associated with the EH and TLS.  In order to estimate Hawking's temperature York conjectured that this irreducible mass corresponds to the mean thermal energy of a quantum oscillator in a heat bath at a given temperature.  The frequencies of oscillation are then determined by the lowest gravitational quasinormal modes of the black hole.  The temperature obtained in this way agrees in good approximation with Hawking's result.\\
York's model provides an example of how it is possible to ``switch on" a quantum ergosphere introducing appropriate quantum effects, namely, quantum oscillations around the classical Schwarzschild metric.  More generally one would expect that the presence of a quantum ergosphere would play a role in the phenomenon of black hole radiance whenever quantum properties of the geometry are taken into account.  Along these lines it is reasonable to assume that quantum effects on the horizon within a particular quantum gravity framework, without the introduction of an {\it ad hoc} model for the quantum fluctuations of the metric, will be effectively described in terms of a quantum ergosphere. In the following section we will see this conjecture at work in the context of the previously discussed tunneling framework.

\section{A tunnel through the quantum horizon}
The emission probability for a shell of energy $E$ put in the form (\ref{Gammaf}) is highly suggestive.  Thinking of the entropy as a measure of the number of micro-states available to a system in a given configuration, the tunneling probability for our shell
\begin{equation}
\label{ampliG}
\Gamma\sim \frac{e^{S_{fin}}}{e^{S_{in}}}=\exp\left(\Delta S\right)\, ,
\end{equation}
is the expression one would expect from a quantum mechanical transition amplitude with a typical dependence on the ratio of the initial and final micro-states of the system given by the entropy change $\Delta S$.\\
This observation calls for an immediate generalization.  Calculations of the black hole entropy in several quantum gravity scenarios \cite{Rovelli:1996dv, Ashtekar:1997yu, Kaul:2000kf, Strominger:1996sh,
Solodukhin:1997yy, Carlip:2000nv, Govindarajan:2001ee, Gour:2002pj}, besides reproducing the familiar linear relation between area and entropy obtained a leading order  ``quantum" correction with a logarithmic\footnote{A similar logarithmic correction to the entropy-area law has also emerged from the calculation of 
one-loop effects of the (quantum) matter fields near the black hole \cite{Fursaev:1994te}. } 
dependence on the area \footnote{We now switch from $k=\hbar=c=G=1$ units of the previous sections to $k=\hbar=c=1$ to keep track of the Planck-scale suppressed terms.}
\begin{equation}
\label{logcorrs}
S_{QG}=\frac{A}{4L_p^2}+\alpha \ln \frac{A}{L_p^2}+O\left(\frac{L_p^2}{A}\right)\,\, ,
\end{equation}
where $\alpha$ is a parameter which depends on the choice of the model.\\  One might expect that a derivation of the emission probability in a quantum gravity framework\footnote{In \cite{Amelino-Camelia:2004xx}, for example, it is shown that the emergence of such a logarithmic correction can be related to Planck-scale modifications of a particle's quantum localization limit.} in presence of back-reaction would lead to an expression analogous to (\ref{ampliG}) with the usual Bekenstein-Hawking entropy $S_{BH}=\frac{A}{4L_p^2}$ replaced by (\ref{logcorrs}) i.e.
\begin{equation}
\Gamma\sim \exp{(S_{QG}(M-E)-S_{QG}(M))}\, .
\end{equation}
The previous expression written in explicit form reads
\begin{equation}
\Gamma(E)\sim \exp{(\Delta S_{QG})}=
\left(1-\frac{E}{M}\right)^{2\alpha}\exp \left(-8\pi GME\left(1-\frac{E}{2M}\right)\right)\, .
\label{prob}
\end{equation}
The exponential in this equation shows the same type of non-thermal deviation found in \cite{Parikh:1999mf}.  In this case, however, an additional factor depending on the ratio of the energy of the emitted quantum and the mass of the black hole is present.  A discussion of the possible consequences of the additional factor for the fate of the black hole in its late stages of evaporation and the information paradox can be found in \cite{Arzano:2005rs}.\\
Our goal here is to show how an emission probability of the type (\ref{prob}) can be obtained if one takes into account the possibility that quantum properties of the background space-time alter the geometry near the horizon.  In the spirit of York we will assume that zero-point quantum fluctuations of the metric produce a splitting between the timelike limit surface and the event horizon.  This would lead to the formation of a quantum ergosphere characterized by the area difference $\delta A_{QE}=\bar{A}_{TLS}-\bar{A}_{EH}$ (where $\bar{A}_{TLS}$ and $\bar{A}_{EH}$ are the mean areas associated with the fluctuating TLS and EH).  As in Section II, in order to derive the tunneling amplitude, we have to evaluate the integral
\begin{equation}
\label{ImS}
\mathrm{Im}\, S=\mathrm{Im}\int_{r_{in}}^{r_{fin}}p_r dr=
\mathrm{Im}\int_{r_{in}}^{r_{fin}}\int_0^H\frac{dH'}{\dot{r}} dr=
-\mathrm{Im}\int_{r_{in}}^{r_{fin}}\int_0^{E}\frac{dE'}{\dot{r}} dr
\,\, ,
\end{equation}
but now taking into account the presence of the quantum ergosphere. Let us focus on the propagation of a classical shell in a Schwarzschild geometry.  When no back reaction effects nor quantum gravity corrections are present the geodesic (\ref{geodesic2}) is simply
\begin{equation}
\dot{r}=\pm1-\sqrt{\frac{2GM}{r}}\, ,
\label{geo1}
\end{equation}
$\dot{r}=0$ at $r=2GM$ where the TLS and EH coincide (the apparent horizon (AH) for spherically symmetric configurations coincides with the TLS)\footnote{The radial coordinate $r$ is, just like in standard Schwarzschild coordinates and in the coordinate set used in \cite{York:1983zb, Bardeen}, the circumferential radius.}. 
To evaluate the effects of this shifting on (\ref{geo1}) we consider the mean irreducible masses associated with the TLS and EH
\begin{equation}
\bar{M}_{TLS}=\left(\frac{\bar{A}_{TLS}}{16\pi}\right)^{1/2},\,\,\,\,\,\,\bar{M}_{EH}=\left(\frac{\bar{A}_{EH}}{16\pi}\right)^{1/2}
\end{equation}
Following \cite{York:1983zb} we assume that $\bar{M}_{TLS}$ and $\bar{M}_{EH}$ will differ from the standard value of $M$ by a term of order $E_p^2/M$
\begin{equation}
\bar{M}_{TLS}=M+\tilde{\alpha}\frac{E_p^2}{M}
\end{equation}
\begin{equation}
\bar{M}_{EH}=M+\tilde{\beta}\frac{E_p^2}{M}
\end{equation}
with $\tilde{\alpha}>\tilde{\beta}$. 
There will be an irreducible mass associated with the quantum ergosphere
$M_{QE}=\bar{M}_{TLS}-\bar{M}_{EH}$ which can be seen as a measure of the zero point energy associated with quantum fluctuations of the geometry.  We assume that a non-vanishing $M_{QE}$ will cause a shift in the pole of the integrand in $\mathrm{Im} S$.  To see this let us recall that, as stressed at the end of section II, the tunneling barrier is set by the energy of the black hole before and after the emission of the shell. This is obtained using only the information about the radial location of the TLS contained in the integral (\ref{imsf}).  We realize then that the position of the TLS is what really determines the emission probability in the tunneling framework.  As an estimate of the shift in the pole we will assume that in the expression for the radial null geodesic (\ref{geo1}) the mass associated with the TLS will be given by the mean value $\bar{M}_{TLS}$. 
Equation (\ref{geo1}) then becomes
\begin{equation}
\label{geo2}
\dot{r}=\pm1-\sqrt{\frac{2G\left(M+\tilde{\alpha}\frac{E_p^2}{M}\right)}{r}}\, 
\end{equation}
As a next step we attempt to introduce back reaction effects due to the energy of the propagating shell.  In doing so let us recall that, in the absence of a quantum ergosphere, a self-gravitating massless shell of energy $E$, in its geodetic motion, ``sees" an effective black hole mass $M-E$, i.e. in the shell's geodesic equation (\ref{geo1})  $M$ is replaced by $M-E$.  Our assumption is that an analogous replacement will be required in (\ref{geo2}) in order to take into account the back reaction of the shell.  The geodesic would then read
\begin{equation}
\label{geo3}
\dot{r}=\pm1-\sqrt{\frac{2G\left((M-E)+\tilde{\alpha}\frac{E_p^2}{(M-E)}\right)}{r}}\, .
\end{equation}
Equipped with this expression we now turn to the calculation of the transition amplitude {\it \`a la} Parikh-Wilczek. Substituting (\ref{geo3}) (with a plus sign for an outgoing shell) in (\ref{ImS}) and integrating over $r$ using the usual Feynman prescription\footnote{The pole is moved in the lower half plane as in \cite{Parikh:1999mf}.} we have
\begin{equation}
\mathrm{Im}\, S=4\pi\int_0^{E} G(M-E')\left(1+\tilde{\alpha}\frac{E_p^2}{(M-E')^2}\right)dE'
\,\, .
\label{Ims}
\end{equation}
Doing the integral over the energy and substituting in (\ref{tunnelampl}) we obtain for the emission probability
\begin{equation}\label{prob2}
\Gamma\sim \exp{(-2\mathrm{Im}S)}=
\left(1-\frac{E}{M}\right)^{8\pi\tilde{\alpha}}\exp \left(-8\pi GME\left(1-\frac{E}{2M}\right)\right)\, 
\end{equation}
which is analogous to (\ref{prob}) provided $\alpha=4\pi\tilde{\alpha}$.

\section{Conclusion}
We adapted the derivation of Parikh and Wilczek in order to include effects due to quantum fluctuations of the horizon.  The ``quantum corrected"  emission probability contains an additional factor 
which depends on the coefficient $\tilde{\alpha}$ which measures the shifting of the TLS from its ``classical" location $r=2M$.  An analogous factor appears in the emission probability when logarithmic corrections to the black hole entropy-area law are present.  This analogy suggests that the quantum ergosphere, seen as an indelible signature of quantum gravity on a black hole metric, affects the near horizon geometry of the black hole leading to the emergence of a logarithmic correction in the entropy-area law.  Reversing the view, the argument we presented might support the idea that leading order (logarithmic) quantum corrections to the black hole entropy are related to the presence of zero-point quantum fluctuations of the metric.\\
\newpage
\begin{center}
{\bf Acknowledgements}
\end{center}
I would like to thank Giovanni Amelino-Camelia for discussions and valuable suggestions, Jack Ng for useful comments and Charles Evans for helpful remarks. I also thank the Department of Physics of the University of Rome "La Sapienza" for hospitality during the course of this work.


\begin{thebibliography}{99}

\bibitem{Bekenstein:1971hc}
J.~D.~Bekenstein,
Phys.\ Rev.\ D {\bf 5}, 1239 (1972).

\bibitem{Hawking:1974sw}
S.~W.~Hawking,
Commun.\ Math.\ Phys.\  {\bf 43}, 199 (1975).

\bibitem{Parikh:1999mf}
M.~K.~Parikh and F.~Wilczek,
Phys.\ Rev.\ Lett.\  {\bf 85}, 5042 (2000)
[arXiv:hep-th/9907001].

\bibitem{Parikh:2004ih}
M.~K.~Parikh,
Int.\ J.\ Mod.\ Phys.\ D {\bf 13}, 2351 (2004)
[arXiv:hep-th/0405160].

\bibitem{Parikh:2004rh}
M.~K.~Parikh,
arXiv:hep-th/0402166.

\bibitem{Massar}
S.~Massar and R.~Parentani,
Nucl.\ Phys.\ B {\bf 575}, 333 (2000)
[arXiv:gr-qc/9903027].

\bibitem{York:1983zb}
J.~W.~York,
Phys.\ Rev.\ D {\bf 28}, 2929 (1983).
  
\bibitem{Rovelli:1996dv}
C.~Rovelli,
Phys.\ Rev.\ Lett.\  {\bf 77}, 3288 (1996)
[arXiv:gr-qc/9603063].

\bibitem{Ashtekar:1997yu}
A.~Ashtekar, J.~Baez, A.~Corichi and K.~Krasnov,
Phys.\ Rev.\ Lett.\  {\bf 80}, 904 (1998)
[arXiv:gr-qc/9710007].

\bibitem{Kaul:2000kf}
R.~K.~Kaul and P.~Majumdar,
Phys.\ Rev.\ Lett.\  {\bf 84}, 5255 (2000)
[arXiv:gr-qc/0002040].

\bibitem{Strominger:1996sh}
A.~Strominger and C.~Vafa,
Phys.\ Lett.\ B {\bf 379}, 99 (1996)
[arXiv:hep-th/9601029]; 

\bibitem{Solodukhin:1997yy}
S.~N.~Solodukhin,
Phys.\ Rev.\ D {\bf 57}, 2410 (1998)
[arXiv:hep-th/9701106].

\bibitem{Carlip:2000nv}
  S.~Carlip,
  Class.\ Quant.\ Grav.\  {\bf 17}, 4175 (2000)
  [arXiv:gr-qc/0005017].
  
\bibitem{Govindarajan:2001ee}
  T.~R.~Govindarajan, R.~K.~Kaul and V.~Suneeta,
  Class.\ Quant.\ Grav.\  {\bf 18}, 2877 (2001)
  [arXiv:gr-qc/0104010].
  
\bibitem{Gour:2002pj}
  G.~Gour,
  Phys.\ Rev.\ D {\bf 66}, 104022 (2002)
  [arXiv:gr-qc/0210024]. 


\bibitem{Fursaev:1994te}
D.~V.~Fursaev,
Phys.\ Rev.\ D {\bf 51}, 5352 (1995)
[arXiv:hep-th/9412161].
  
\bibitem{Amelino-Camelia:2004xx}
G.~Amelino-Camelia, M.~Arzano and A.~Procaccini,
Phys.\ Rev.\ D {\bf 70}, 107501 (2004)
[arXiv:gr-qc/0405084],  
G.~Amelino-Camelia, A.~Procaccini and M.~Arzano,
Int.\ J.\ Mod.\ Phys.\ D {\bf 13}, 2337 (2004).
G.~Amelino-Camelia, M.~Arzano, Y.~Ling and G.~Mandanici,
  arXiv:gr-qc/0506110.


\bibitem{Kraus:1994by}
P.~Kraus and F.~Wilczek,
Nucl.\ Phys.\ B {\bf 433}, 403 (1995)
[arXiv:gr-qc/9408003].

\bibitem{Keski-Vakkuri:1996xp}
E.~Keski-Vakkuri and P.~Kraus,
Nucl.\ Phys.\ B {\bf 491}, 249 (1997)
[arXiv:hep-th/9610045].


\bibitem{Kraus:1994fh}
P.~Kraus and F.~Wilczek,
arXiv:gr-qc/9406042.

\bibitem{Bardeen}
J.~M.~Bardeen,
Phys.\ Rev.\ Lett.\  {\bf 46}, 382 (1981).

\bibitem{York:1983fx}
  J.~W.~York,
in {\it Quantum Theory Of Gravity}, 135-147 edited by S. Christensen.

\bibitem{Arzano:2005rs}
  M.~Arzano, A.~J.~M.~Medved and E.~C.~Vagenas,
  JHEP {\bf 0509}, 037 (2005)
  [arXiv:hep-th/0505266].

\end{thebibliography}
\end{document}